\documentclass[sigconf]{acmart}

\usepackage{algorithm}
\usepackage{algpseudocode}
\usepackage{framed}
\usepackage{listings}
\lstdefinestyle{promptstyle}{
    basicstyle=\ttfamily\small,
    breaklines=true,
    breakatwhitespace=true,
    frame=none
}

\AtBeginDocument{%
  }

\setcopyright{acmlicensed}
\copyrightyear{2026}
\acmYear{2026}
\setcopyright{cc}
\setcctype{by}
\acmConference[UIST '26]{The 39th Annual ACM Symposium on User Interface Software and Technology}{November 02--05, 2026}{Detroit, MI, USA}
\acmBooktitle{The 39th Annual ACM Symposium on User Interface Software and Technology (UIST '26), November 02--05, 2026, Detroit, MI, USA}
\acmDOI{10.1145/3830398.3830532}
\acmISBN{979-8-4007-2856-3/2026/11}

\begin{document}

\title{Before You Say It: Anticipating Verbal Behavior from Longitudinal Everyday Conversations with LLMs}





\author{Yasith Samaradivakara}
\authornote{Both authors contributed equally to this research.}
\email{yasith@mit.edu}
\affiliation{%
  \institution{Massachusetts Institute of Technology}
  \city{Cambridge}
  \state{MA}
  \country{USA}
}

\author{Valdemar Danry}
\email{vdanry@mit.edu}
\authornotemark[1]
\affiliation{%
  \institution{Massachusetts Institute of Technology}
  \city{Cambridge}
  \state{MA}
  \country{USA}
}

\author{Paul Liang}
\affiliation{%
  \institution{Massachusetts Institute of Technology}
  \city{Cambridge}
  \state{MA}
  \country{USA}
}

\author{Pattie Maes}
\affiliation{%
  \institution{Massachusetts Institute of Technology}
  \city{Cambridge}
  \state{MA}
  \country{USA}
}

\renewcommand{\shortauthors}{Samaradivakara, Danry, et al.}

\begin{abstract}
Knowing someone deeply means not just understanding what they say or do but also how they will likely think, react, and engage across situations. Such predictions could eventually inform systems to anticipate when the individual is about to deviate from their goal, catch regrettable behaviors before they are made, and surface blind spots before they take hold. While many interactive systems model users to enable more personalized interactions, most cannot make such behavioral predictions, as this often requires longitudinal observation and inference of how the individual's behaviors unfold across various everyday situations. In this work, we introduce a novel LLM-based predictive behavioral modeling approach that anticipates a user's likely behavior across everyday conversational situations. We (1) collect a longitudinal dataset of over 1,000 hours of naturalistic conversations from 14 participants using a wearable smartwatch; (2) evaluate LLM-based predictions against ground truth behaviors; and (3) use semi-structured interviews to explore participants’ perceptions of behavioral predictions and their views on possible forms of future behavioral support. Altogether, our findings provide evidence that person-specific verbal behavior can be predicted from longitudinal conversational data. This opens up new possibilities for potential future context-aware, anticipatory, proactive and personalized AI systems.



\end{abstract}


\begin{teaserfigure}
  \includegraphics[width=\textwidth]{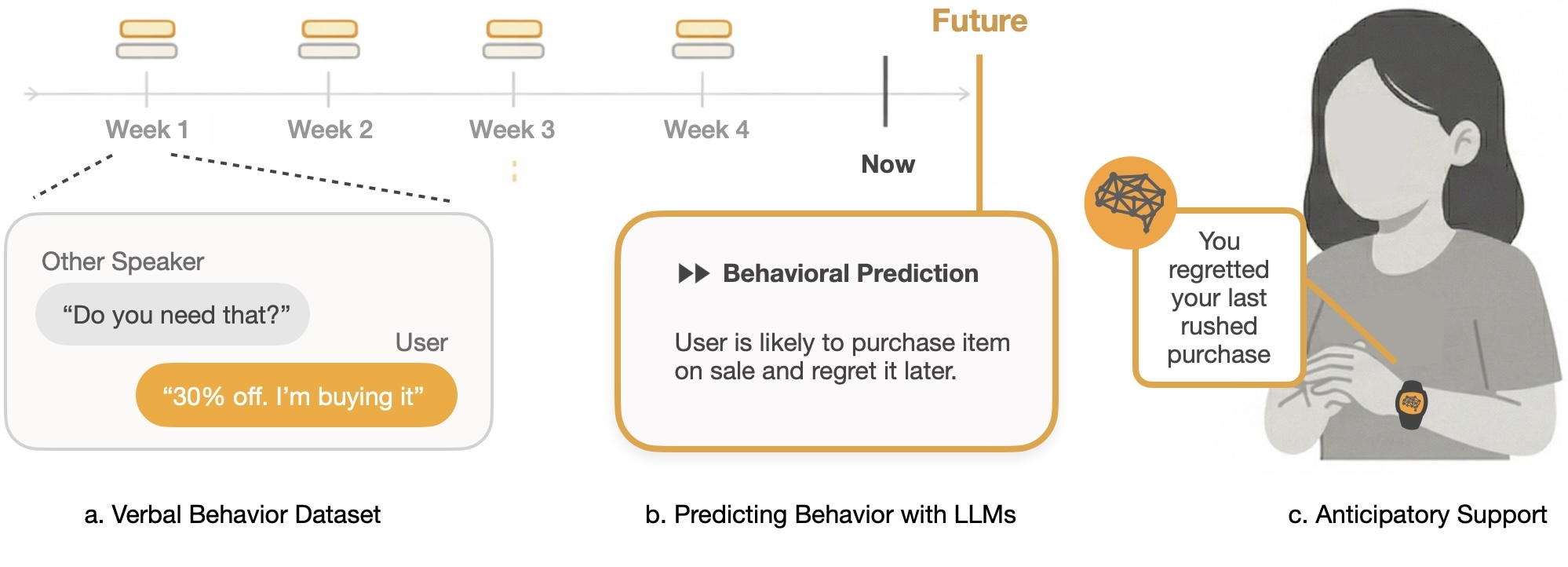}
  \caption{Conversational data being used to mine behavioral patterns of a user and predict their likely next behavior. First, longitudinal conversational data reveals a recurring impulse purchasing pattern the user wants to change. Our prediction approach identifies when a similar conversational situation may be emerging. In the future, a system might use behavioral predictions to surface a just-in-time personalized reminders before the user's action unfolds.}
  \Description{Enjoying the baseball game from the third-base
  seats. Ichiro Suzuki preparing to bat.}
  \label{fig:teaser}
\end{teaserfigure}




\maketitle

\section{Introduction}

People who know us well can often tell what we are about to do before we do it. A close friend can hear the first few sentences of a conversation and sense that we are about to pick the same fight again. A partner can notice when we are rationalizing a purchase we will later regret. This kind of anticipation is valuable because it enables others to helps us \emph{before} something happens: to surface blind spots, warn us when we are drifting from our intentions, and create a critical small window in which we can choose differently. This ability arises from learning person-specific regularities that link situations to behavior, similar to how behavior is identified in decades of psychological research \cite{mischel1995cognitive, tversky1974judgment, habits}.

Despite progress in AI assistants that answer questions, retrieve information, and respond to explicit requests, they remain largely reactive~\cite{shaikh2025creating,deng2024towards}.
An assistant that learns a user's recurring conversational tendencies could anticipate likely next moves and intervene at moments that matter.
With recent advancements, wearable devices can now capture conversational traces in everyday life \cite{rhodes1997remembrance, zulfikar2024memoro,maniar2025mempal}, and LLMs can reason over naturalistic histories to build richer user representations 
\cite{park2024generative, shaikh2025creating}. Yet a gap remains in turning 
longitudinal everyday conversation into a predictive model of a user's \emph{situation-dependent verbal behavior}.


In this paper, we extend behavioral pattern mining approaches introduced in previous work \cite{danry2026mind} with Pattern-Conditioned Prediction, an interpretable method that uses behavioral patterns mined from longitudinal everyday conversations to anticipate verbal behavior in new conversational contexts. Using 1{,}000+ hours of naturalistic speech collected from 14 participants wearing an always-on smartwatch, we study whether LLMs can predict how a person is likely to respond next in conversational situations. We frame this as a problem of verbal behavioral prediction—predicting the communicative intention of a user’s next response rather than surface utterances—grounded in longitudinal personal history, with the goal of enabling interactive systems to anticipate conversational trajectories before they unfold. The paper makes the following contributions:

\begin{itemize}
    \item We contribute a longitudinal wearable conversation dataset for studying verbal behavior in everyday life, comprising over 1{,}000 hours of naturalistic speech from 14 participants.
    \item We introduce \emph{Situational Reasoning}, 
    a training-free approach that learns 
    interpretable, situation-dependent behavioral 
    patterns as they unfold over time, and show 
    it significantly outperforms existing LLM baselines when predicting verbal behavior from unfolding conversational context.
    \item We examine prediction on behaviors participants identified as ones would like to change or get support with, and explore through semi-structured interviews how behavioral anticipation could inform the design of proactive interventions.
\end{itemize}

\section{Related Work}
Work on ambulatory and wearable sensing has established the feasibility of capturing conversation in everyday life. Early systems such as the Electronically Activated Recorder (EAR) demonstrated that lightweight devices can sample naturalistic audio outside laboratory settings \cite{mehl2001electronically} and support memory and reflection \cite{vemuri2004audio, vemuri2006iremember}. Broader mobile and smartphone sensing research has since established methods for collecting longitudinal behavioral data unobtrusively in everyday contexts, including social interactions, activity patterns, and conversational 
behavior \cite{harari2016using, harari2023understanding}. More recent wearable systems combine continuous conversational sensing with large language models to support memory augmentation, retrieval, reflection, and behavioral modeling in daily life \cite{zulfikar2024memoro,maniar2025mempal,danry2026mind}. While these approaches make it increasingly practical to collect and process longitudinal conversational traces, they primarily focus on capturing, summarizing, retrieving, or reflecting on past interactions, rather than modeling how an individual is likely to behave in a specific unfolding conversational context.

A second line of work studies communicative function and personalization in language systems. Verbal Response Modes (VRM) and related dialogue-act frameworks provide structured representations of communicative intent, and have been widely used to classify utterances or predict conversational acts from local context \cite{stiles1992describing,stolcke2000dialogue}. In parallel, personalized dialogue systems condition generation on user profiles, inferred personas, or long-term interaction histories \cite{zhang2018personalizing}, while recent LLM-based assistants increasingly maintain persistent memory across sessions \cite{chen2024large,shaikh2025creating,deng2024towards} and user modeling for recommendation and personalized text generation \cite{salemi2023lamp, ning2024user}. These approaches typically aim to improve assistant response generation or model generic conversational structure, and do not explicitly capture person-specific, situation-dependent patterns in how individuals behave across diverse real-world interactions.

Finally, prior work on proactive assistants and just-in-time adaptive interventions highlights the value of anticipating user needs before they are explicitly expressed \cite{berube2024proactive,nahum2016just, nahum2015building, satori, sensibleagent}. Applications discussed in this literature span communication support, behavior change interventions, task automation, and context-aware reminders, where timely assistance can reduce friction, prevent errors, or support better decisions \cite{berube2024proactive,oh2024better, mirai}. Yet these systems largely rely on immediate context, predefined heuristics, or population-level patterns, rather than predictions grounded in a user’s longitudinal behavioral history. As a result, existing approaches remain limited in their ability to anticipate how a particular individual is likely to act in a concrete situation as it unfolds. This paper addresses this gap by modeling person-specific verbal behavior from longitudinal everyday conversations and using it to predict likely communicative actions in context.

\section{Longitudinal Wearable Data Collection}

\subsubsection*{Participants and Procedure}
We recruited 14 participants (Age: 18+, fluent English speakers, no speech impairments) who each wore an always-on smartwatch for 7--10 days during normal daily activities, compensated \$100 USD each. The watch ran a background application using on-device voice activity detection, forwarding 2--3 minute audio clips to a backend pipeline when speech was detected.  Participants were instructed to inform others around them that conversations were being recorded and to obtain their consent prior to each interaction. The study was approved by the institutional review board MIT COUHES \#2402001229.

\subsubsection*{Audio Processing and Transcription}
Audio was processed through a backend pipeline that (1) transcribed and diarized speech using Deepgram's \texttt{nova-3-meeting} model, (2) identified the target participant's voice using a speaker verification model trained on a 20-second enrollment sample provided at onboarding, tagging the participant's utterances as ``User'' or ``Other'', and (3) anonymized personally identifiable information using SpaCy's Named Entity Recognition model (e.g., [PERSON], [ORGANIZATION]). Raw audio was deleted immediately after processing and transcripts were stored locally on the device using AES-256-GCM encryption.

\subsubsection*{Data Review \& Statistics}
Following the data collection period, participants reviewed their transcribed conversations via a web-based interface, removing data they did not wish to share and correcting misattributed speaker labels and conversational context (e.g. role, situation, emotion etc.). On average, they removed 0.16\% of transcribed data and flagged 2.48\% of segments as misclassified. The final dataset comprises 15,066 utterances across 14 participants (mean: 1,256 per participant), of which 57\% as ``User'' and 43\% as ``Other'', with a mean utterance length of 49 words. Further dataset statistics are provided in appendix \ref{corpus}.

\subsubsection*{Data Cleaning}
We further applied three data cleaning steps to enhance data quality: (1) \textbf{Word count filtering}: turns where the user utterance or the interlocutor turn contained fewer than 4 words were excluded, removing trivial exchanges that provide insufficient content for meaningful evaluation. (2) \textbf{Disfluency removal}: filler words (e.g., \textit{um}, \textit{uh}, \textit{like}), false starts, and repetitions common in transcribed naturalistic speech were detected using a BERT-based disfluency classifier \cite{hf-bert-disfluceny}.(3) \textbf{Utterance re-segmentation}: sentence boundary detection was applied using \texttt{wtpsplit}~\cite{minixhofer-etal-2023-wheres} to merge fragmented units into coherent utterance-level turns. After cleaning, 9,901 utterances were retained.

\section{Predictive Behavior Modeling}
\subsubsection{Task} We first define the prediction task. In Verbal Response Mode (VRM) theory~\cite{stiles1992describing,stolcke2000dialogue} verbal behavior refers to the communicative intention an utterance serves in a conversation independent of its surface linguistic form. Following this, we define verbal behavior prediction task as follows.

A conversation $\mathcal{U} = \{(u_i, s_i)\}_{i=1}^{T}$ is a sequence of utterance-speaker pairs, where $s_i \in \{\mathcal{S}_u, \mathcal{S}_o\}$ denotes whether the speaker is the target user or one or more other speakers, and each conversation is bounded by a distinct interaction episode. Given a target user turn $t$, the conversational 
context $\mathcal{C}_t = \{(u_i, s_i)\}_{i=1}^{t-1}$ comprises all preceding utterances in that episode.


We formulate verbal behavior prediction as generating a predicted verbal behavioral tendency $\hat{B}_t$, a natural language characterization of the communicative intention or function that the user $\mathcal{S}_u$ is likely to express next. This generative formulation produces generalized, actionable behavioral predictions (for example, \textit{``the user is  likely to minimize their distress when asked directly''}) rather than direct intent \textit{``the user will disclose''}, and is considered at two levels of granularity: single-utterance, where $\hat{B}_t$ describes 
a single user turn, and multi-utterance, where it describes communicative function across a window of consecutive user turns.

We instantiate this task using Large Language Models (LLMs) based on \textit{Gemini 2.5 Pro} \cite{gemini25pro_card}, prompted with the VRM taxonomy to generate $\hat{B}_t$ given  $\mathcal{C}_t$ (see Appendix \ref{appx:prediction_prompts} for full prompts). We define two comparison conditions for this task following existing methods \cite{park2024generative}, and introduce a new method 'Situational Reasoning'. 


\subsubsection*{Baseline 1: Zero-Shot} This represents the simplest possible instantiation of the task, with no personalization or longitudinal modeling. It generates $\hat{B}_t$ from the local conversational context alone, with no access to prior interactions,  $\hat{B}_t = f(\mathcal{C}_t)$. 

\subsubsection*{Baseline 2: All-In-Context} The all-in-context baseline extends this by providing the full longitudinal history $\mathcal{H}$ of prior conversations, conditioning prediction on both history and local context,  $\hat{B}_t = f(\mathcal{H}, \mathcal{C}_t)$, operationalizing behavioral prediction as long in-context learning \cite{brown2020gpt3} where the model implicitly infers person-specific tendencies from the accumulated transcript. History is truncated to the most recent $500{,}000$ characters when it exceeds the model's context window.

\subsubsection*{Baseline 3: Natural-Language Summary}
To isolate the contribution of the proposed representation from information distillation alone, we introduce a natural-language summary baseline. The model summarizes the user's longitudinal interaction history into a concise narrative of recurring behavioral tendencies in free-form natural language, following prior work on LLM-based summarization \cite{shaikh2025creating, brown2020gpt3}. The generated summary replaces the full history and is provided with the current conversational context to predict the user's next verbal behavioral tendency (see Appendix \ref{appx:prediction_prompts} for prompts).

\begin{figure*}
    \centering
    \includegraphics[width=1\linewidth, height=5cm]{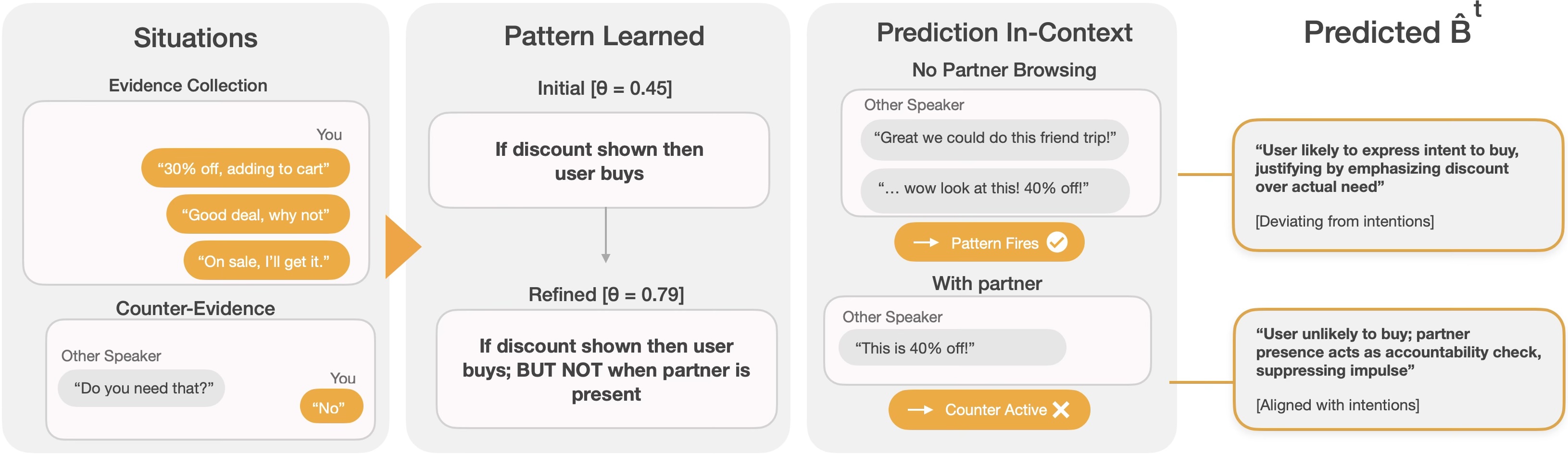}
    \caption{Four stages of pipeline: (1) pattern mining, evidence  and counter-evidence collection from longitudinal 
    conversations, (2) pattern induction and 
    refinement with confidence scores, (3) 
    pattern activation based on the current context, 
    and (4) generation of behavioral predictions 
    reflecting users likely behavior}
    \label{fig:system}
\end{figure*}

\subsection{Situational Reasoning}
Decades of psychological research show that human behavior follows stable, situation-specific regularities rather than global traits \cite{mischel1995cognitive}, presenting a promising basis for learning behavioral models across situations. While recent advances in LLMs offer the capacity to reason over unstructured data to infer latent patterns across situations \cite{brown2020language, wei2022chain, shaikh2025creating, generativeagents}, conditioning generation directly on raw longitudinal history is brittle: models struggle to surface person-specific regularities buried in long contexts \cite{liu2024lost}, and outputs are difficult for users to inspect or correct \cite{ji2023survey}. Symbolic representations such as ``IF X, THEN Y''-rules address this by converting implicit behavioral patterns into explicit, interpretable rules that guide generation precisely, naturally reflecting the situation-specific structure of human behavior making it predictable \cite{mischel1995cognitive, danry2026mind}. 

Extending this work, we mine situation-dependent behavioral patterns from longitudinal conversations and design a novel predictive modeling approach that learns \emph{situational behavioral patterns} from longitudinal conversation and uses them to guide behavioral prediction in context at inference time. 
\paragraph{Behavioral Patterns.}
We define each pattern as a \emph{contrastive IF-THEN-EXCEPT} rule: 
\begin{quote}
\small
IF [SITUATION] $\phi$, THEN behavior $\psi$, \\
BUT NOT when [EXCEPTION-SITUATION] $R$.
\end{quote}

where the behavioral response $\psi_k$ is associated with the situational antecedent $\phi_k$, unless the exception condition $R_k$ is present. Each pattern consists of (1) a \textbf{situational condition}, describing \emph{when} (time or activity), \emph{where} (setting), \emph{with whom} (participant role, relationship, conversational move), and \emph{internal state} (e.g., emotion) under which a behavior tends to occur \cite{omniquery}; (2) a \textbf{counter-situation}, specifying when an otherwise applicable tendency should be suppressed; and (3) a \textbf{behavioral tendency}, describing the user's likely communicative response. For example,
\begin{quote}
\textit{IF challenged by an authority figure, THEN the user tends to deflect, BUT NOT when challenged by a peer.}
\end{quote}

\paragraph{Behavioral Pattern Mining.}
Each user's behavioral profile is learned incrementally from longitudinal conversations by identifying recurring associations between situations and behaviors. 
Motivated by prior work on inductive rule learning \cite{yang2022language, Park2023GenerativeAgents}, patterns with overlapping conditions are merged into more general patterns as evidence accumulates. Pattern probability is then estimated from the relative balance of supporting and contradicting instances:
\[
\theta_k \propto \frac{|E_k|}{|E_k| + 
|\bar{E}_k|} \in [0,1],
\]
and mapped to probabilistic qualifiers — \emph{never}, \emph{sometimes}, \emph{often}, \emph{always} — following established approaches of probabilistic judgments from language models \cite{wang2025always}. Patterns are updated as new interactions arrive, allowing their evidence, confidence, and estimated probability to be refined over time.






\paragraph{Pattern Activation and Prediction} While behavioral pattern mining builds the behavioral model as interactions unfold, prediction requires determining which patterns apply to the current conversational context. Given the current conversational context $x_{t-1}$, each pattern is evaluated by matching its situational antecedent $\phi_k$ and checking that its exception condition $R_k$ is absent. The activated patterns, weighted by their estimated probability $\theta_k$ are provided to an LLM alongside the current conversational context to generate a prediction of the user’s next verbal behavior. An example of the pipeline is shown in figure \ref{fig:system}.

\begin{figure*}
    \centering
    \includegraphics[width=1\linewidth]{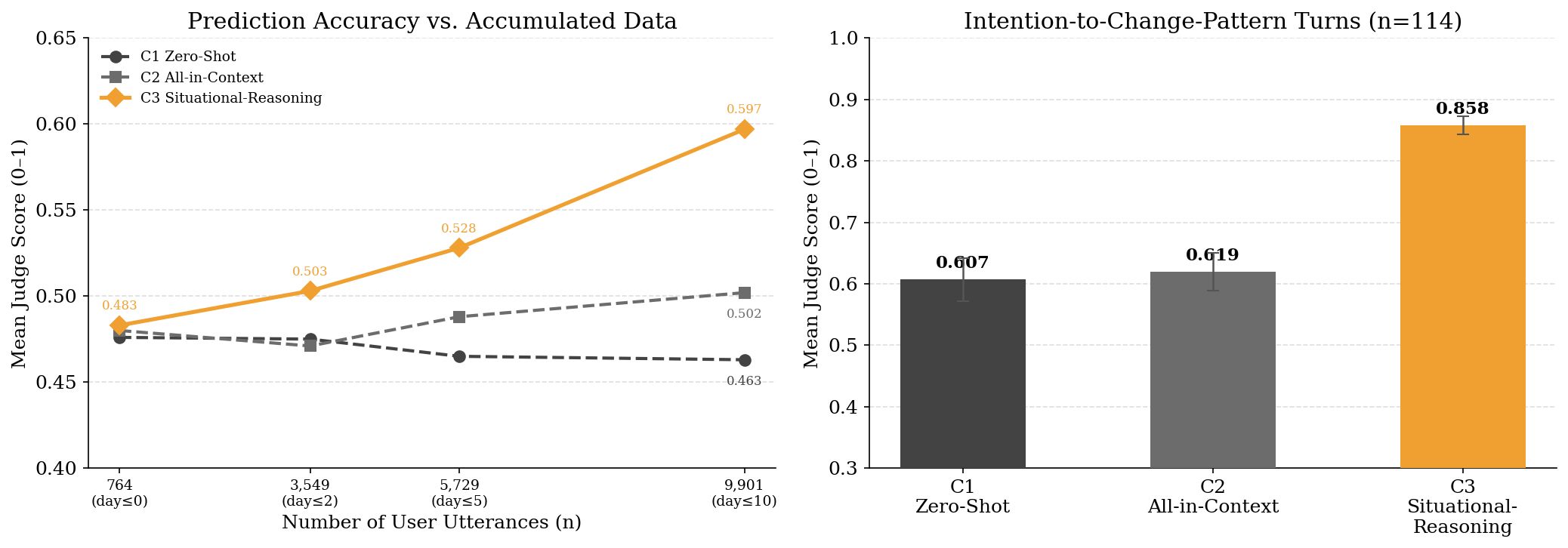}
    \caption{(a) Prediction accuracy as a 
    function of accumulated conversational data. 
    Situational Reasoning improves consistently 
    with more data, reaching 0.597 at full 
    accumulation, while Zero-Shot and 
    All-in-Context remain relatively flat. 
    (b) Mean scores on participant-flagged 
    intention-to-change patterns, where 
    Situational Reasoning substantially 
    outperforms both baselines.}
    \label{fig:overall}
\end{figure*}

\section{Evaluation}
We evaluated the four prediction methods (Zero Shot, All-in-Context, NL-Summaries, Pattern-Conditioned Prediction) by comparing the predicted behavioral description $\hat{B}_t$ against the ground-truth behavioral description $B_t$ with the conversational context, generated from the actual utterance using the same prompt. Predictions were assessed in the context of the preceding conversation using both LLM-based evaluation and human annotations grounded in communicative intention alignment from the VRM taxonomy. We further validated the LLM judgments against 40 independent human evaluators to establish inter-rater reliability.

\subsubsection*{LLM-as-a-Judge Evaluation}
We evaluate predictions along three dimensions using automated LLM-as-a-judge scoring \cite{zheng2023judging}, each scored on $[0, 1]$: (1) \textbf{Pragmatic Function Alignment} ($d_1$), whether the prediction captures the same verbal intention as the ground truth independent of surface wording. For example, predicting \textit{``the user is likely to deflect''}, when the ground truth reflects withdrawal, scores higher than predicting \textit{``the user will ask a clarifying question''}, grounded in VRM functional categories \cite{stiles1992describing}; (2) \textbf{Behavioral Specificity} ($d_2$), whether the prediction matches the level of specificity of the ground truth rather than producing a more generic description — for example, \textit{``the user minimizes distress and redirects to logistics when overwhelmed''} scores higher than \textit{``the user may express concern''} \cite{vallacher1987action, qi2020stay}; and (3) \textbf{Compositional Alignment} ($d_3$), whether the prediction captures multiple communicative functions when present and reflects their relative weight appropriately \cite{stiles1992describing}. We use GPT-5\footnote{\url{https://openai.com/index/gpt-5-system-card/}} as the judge (Appendix~\ref{appx:judge_prompt} for Prompts).

\subsection{Prediction Results}
\subsubsection*{LLM Prediction Results} Our results show that LLMs can meaningfully predict verbal behavior. The LLM-judge score of pattern conditioned prediction \textbf{M: 0.597, SD: 0.350}, significantly outperforms both the zero-shot baseline by \textbf{+28.9\%} (Zero-Shot: M-0.463, SD-0.330) and the all-in-context retrieval condition by \textbf{+18.9\%} (All-In-Context: M-0.502, SD-0.339). Accumulated performance over time and data size is increasingly improved for pattern-conditioned prediction (+23.6\%) while for zero-shot and all-in-context remains flat, Figure~\ref{fig:overall}. Per participant results,  activation rates, and example predictions are shown in Appendix ~\ref{appx:examples}. To further evaluate whether predictions are person-specific rather than generally accurate, we evaluated a cross-participant transfer condition (Pattern-conditioned prediction-cross), in which each participant's behavioral patterns are replaced by those of a randomly assigned other participant. Pattern-conditioned prediction-cross scores \textbf{M-0.460, SD-0.389} compared to pattern-conditioned prediction by \textbf{+29.8\%}. 

\subsubsection*{Human Evaluation Results}
To assess the reliability of the LLM-as-a-judge evaluation, we conducted a human evaluation with 40 independent crowdsourced raters on a randomly sampled set of 200 prediction scenarios. Raters compared the four prediction methods (Zero-Shot, All-in-Context, Natural-Language Summary, and Pattern-Conditioned Prediction) using the same evaluation protocol as the authors and LLM judge. They ranked overall prediction quality and evaluated Behavioral Specificity ($d_2$) and Compositional Alignment ($d_3$), while Pragmatic Function Alignment ($d_1$) was assessed using the VRM taxonomy. Pattern-Conditioned Prediction ranked first in 43\% of comparisons (vs. All-in-Context: 24\%, Natural-Language Summary: 15\%, and Zero-Shot: 18\%), with strong agreement among raters (Kendall's $\tau = 0.83$). The human rankings closely matched both the author annotations and the LLM-as-a-judge evaluation, supporting the reliability of the automated evaluation.

\subsection{Semi-Structured Interviews}
To understand how participants engaged with their inferred behavioral patterns and how behavioral prediction could support real-world behavior change, we conducted two studies. First, all participants reviewed a subset of inferred patterns and flagged reflecting behaviors they wanted to change. Second, a subset of participants returned for follow-up interviews several months later to reflect on those patterns, discuss strategies they had tried, and envision what forms of interventions would be most useful.

\subsubsection*{Pattern Review}
All participants in total flagged 114 (Mean: 12) patterns detected by our behavioral pattern extraction method as reflecting behaviors they wanted to change (\textit{"No", "Yes"}). Each pattern had an average of 8 activations in their conversation history predicted by our method. On this subset ($n=912$), pattern-conditioned prediction achieves a score of \textbf{0.858}, significantly outperforming the zero-shot baseline by +41.3\% (0.607) and all-in-context condition by +38.5\% (0.619). This suggests that pattern-conditioned prediction is particularly effective for recurring behaviors that participants personally recognized and wanted to change.

\subsubsection*{Follow-up Discussions}
Few months after initial pattern review, we conducted follow-up semi-structured interviews with 7 participants, revisiting the patterns they had previously flagged as behaviors they wanted to change. For several participants, simply seeing these patterns appeared to support \textbf{reflection and greater self-awareness}, sometimes helping them begin to change their behavior (P11, P13, P15, P10). Participants described \textbf{three recurring challenges} in changing unwanted tendencies: \textbf{(1) falling back on default habits under stress or when deeply ingrained} (P11, P15, P6), \textbf{(2) uncertainty in socially ambiguous situations or around authority figures} (P13, P5), and \textbf{(3) difficulty noticing the pattern early enough in the moment} (P6, P11, P15). They also described strategies that help overcome these challenges, suggesting design opportunities for anticipatory AI systems. Some participants emphasized \textbf{private, nonjudgmental reminders} that prompt reflection without shame, preferring machine-delivered nudges over interpersonal correction; as P11 put it, a watch-based nudge would \textit{``less judgmental''} and \textit{``more like information,''} while P15 valued reminders because they are \textit{``something I can't really avoid.''} (P11, P15, P6). 
Others described benefits from \textbf{reflective self-checking, rehearsing difficult conversations, or pausing to reevaluate priorities before responding} (P13, P15). Participants also highlighted that interventions should be \textbf{situationally tailored}: suggestions to reframe the moment, redirect attention, offer alternatives, reduce commitment size, or reinforce previously stated goals were seen as more helpful than generic prompts (P10, P11, P5). 
At the same time, some participants were cautious about \textbf{mistimed or overly directive interventions}, especially in nuanced “gray area” situations where the system might miss context or undermine personal agency (P13). Together, these findings suggest that anticipatory AI for behavioral support may be most useful when it surfaces human-readable patterns for reflection and offers private, situationally appropriate forms of support aligned with strategies users already find effective in everyday life.

\section{Discussion \& Future Work}
Our results show that situation-specific behavioral patterns mined from longitudinal conversations can meaningfully improve predictions of users’ verbal behavioral tendencies. A key strength of our approach is representing these as human-readable context-conditioned behavioral patterns with explicit exception conditions that users can inspect and refine. Participants valued this transparency, with P10 appreciating being able to read, correct, and set personal goals for their inferred patterns. Participants also reflected on their own strategies for changing behavior, suggesting how proactive systems could help by redirecting attention, offering alternatives, and surfacing reframings tied to their goals. This points toward anticipatory AI that works \emph{with} users — grounding interventions in patterns they recognize and strategies they already find effective. Extending this framework through reinforcement learning from user feedback could enable adaptive refinement over time, while applying it to longer-horizon user modeling across messaging, social media, and conversational platforms. With these promising results, several limitations remain. Data quality is constrained by microphone noise and a limited deployment window. Also, predictions were evaluated at the turn level; longer scales and cross-context generalization remain open. Our predictions were also limited by the availability of activated situational behavioral patterns, motivating future work on more robust fallback mechanisms when personalized situational evidence is sparse or unavailable. More broadly, our findings suggest a path toward wearable systems that anticipate users’ future verbal behavior from unfolding conversational context, supporting the design of proactive, context-aware human–AI interactions.

\bibliographystyle{ACM-Reference-Format}
\bibliography{ref}

\appendix

\section{Additional Results}

\begin{figure*}
    \centering
    \includegraphics[width=1\linewidth]{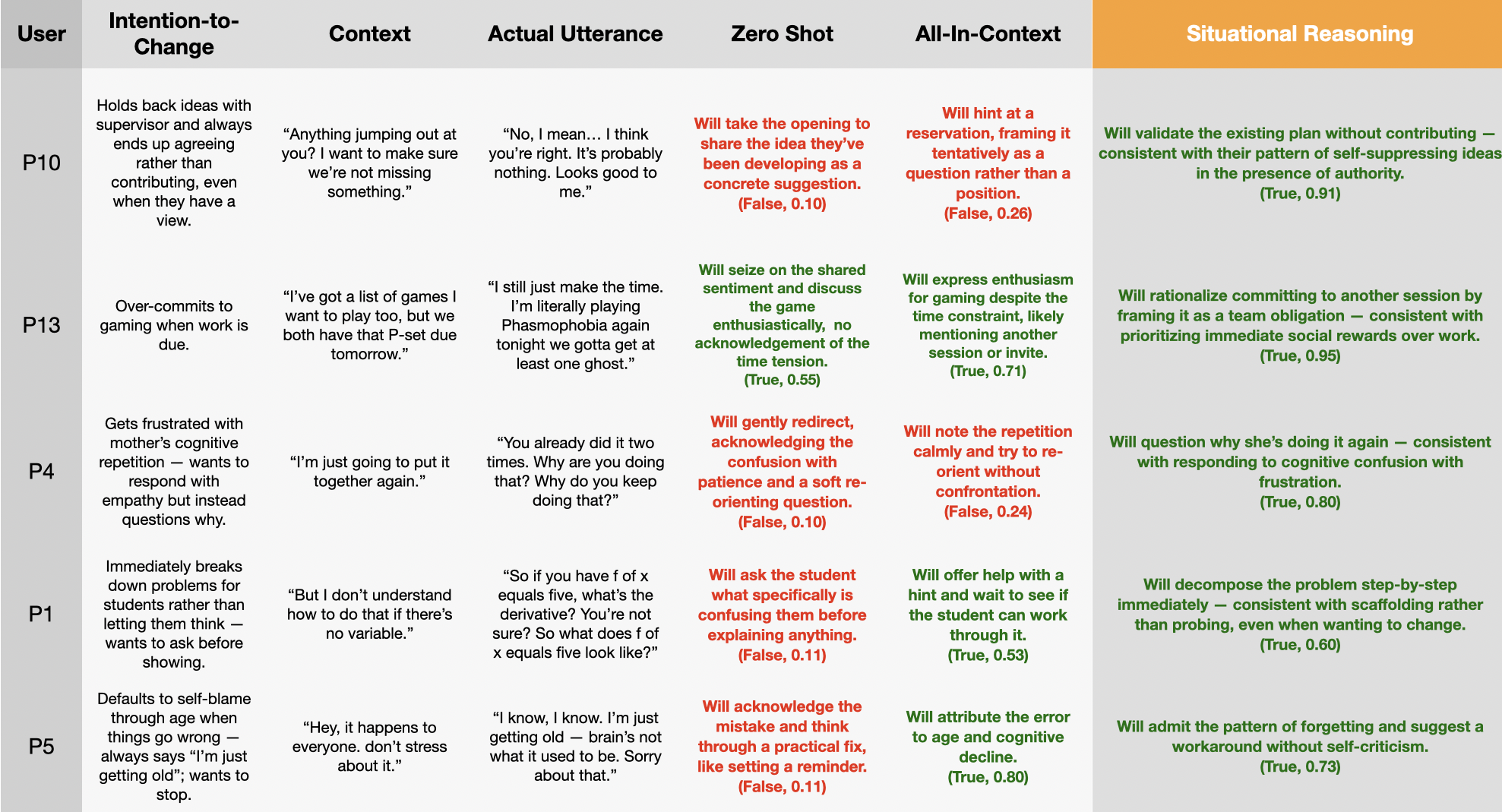}
    \caption{Comparison of prediction methods across real-world conversational scenarios. Pattern-conditioned prediction more accurately captures user-specific behavioral tendencies, particularly in cases where users express an intention to change recurring patterns. Utterances are lightly edited for readability without altering their original meaning.}
    \label{appx:examples}
\end{figure*}

\begin{figure*}
    \centering
    \includegraphics[width=1\linewidth]{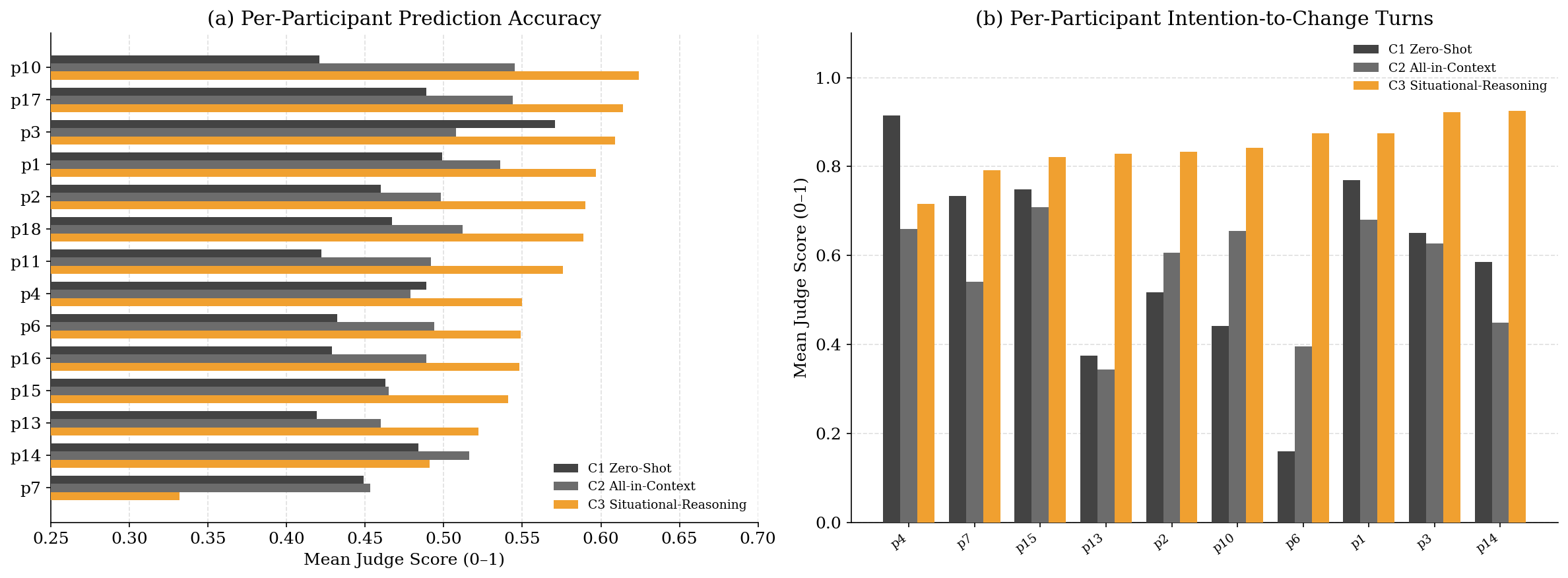}
    \caption{Per-participant breakdown of prediction scores and intention-to-change flagged moments. (a) Mean judge scores per participant across all conversational turns, sorted by C3 performance. C3 outperforms C1 and C2 for most participants, with the exception of p7 where C3 degrades, suggesting that the quality of mined patterns via inductive inference varies across participants. (b) Mean judge scores per participant restricted to intention-to-change turns, where C3 shows larger and more consistent gains over C1 and C2, indicating that pattern-conditioned prediction is especially effective on turns where the underlying behavioral tendency has personal salience for the participant.}
    \label{appx:participant}
\end{figure*}

\begin{figure*}
    \centering
    \includegraphics[height=10cm]{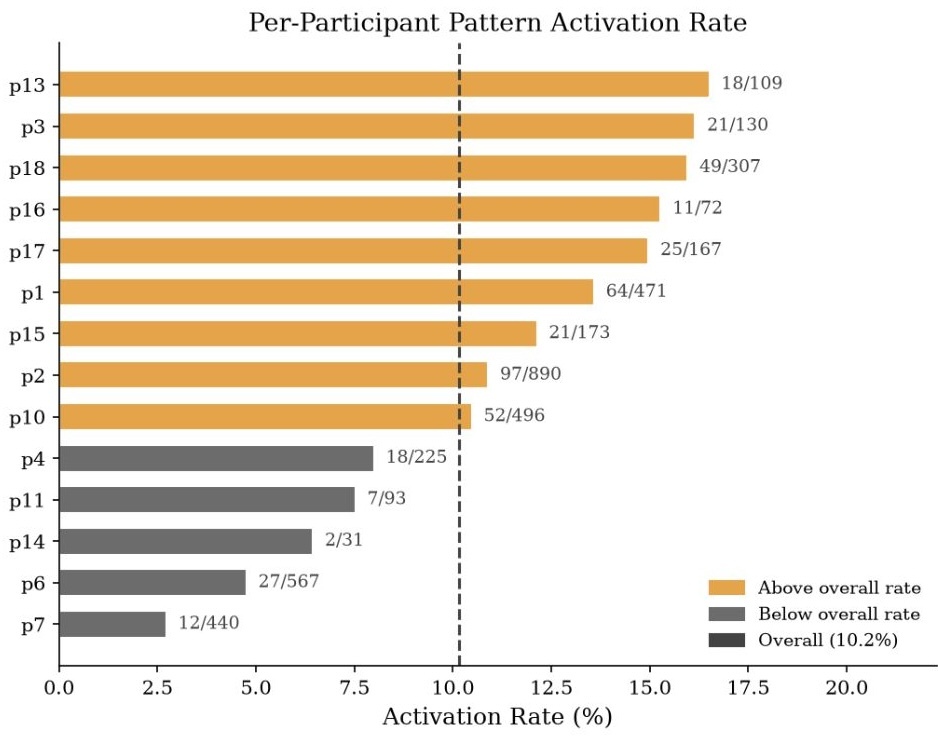}
    \caption{Per-participant pattern activation rates during inference. Numbers indicate the number of activated behavioral patterns relative to the total prediction instances. Prediction performance depends on the availability of activated situational evidence; participants with lower activation rates rely more frequently on the fallback prediction mechanism when no relevant behavioral pattern is activated (e.g., P7), contributing to reduced performance}
    \label{fig:desc}    
    \Description{Per-participant pattern activation rates during inference. Numbers indicate the number of activated behavioral patterns relative to the total prediction instances. Prediction performance depends on the availability of activated situational evidence; participants with lower activation rates rely more frequently on the fallback prediction mechanism when no relevant behavioral pattern is activated (e.g., P7), contributing to reduced performance}
\label{corpus}
\end{figure*}

\section{Prompts}
\subsection{Prediction Prompts}
\label{appx:prediction_prompts}

\subsubsection*{C1: Zero-Shot Baseline}

\begin{framed}
\begin{lstlisting}[style=promptstyle]
You are analyzing a real person's verbal 
behavior in a conversation. The transcript 
below captures a real conversation with this 
specific individual.

Your task is NOT to predict the exact words 
they will say. Instead, predict their verbal 
behavioral tendency -- the communicative 
intention and function their next utterance 
is most likely to serve, grounded in VRM 
theory.

{VRM_TAXONOMY}

When forming your prediction, consider:
- Which VRM function their next turn will 
  most likely perform (e.g., Disclosure, 
  Advisement, Confirmation, Question)
- The emotional and relational register they 
  are likely to bring (e.g., hesitant, 
  collaborative, defensive)
- How they have been orienting in this 
  conversation so far (e.g., reflective, 
  avoidant, problem-focused)

Conversation so far:
{context_text}

Respond with:
Behavioral tendency: <2-3 sentences 
describing the verbal behavior [User] will 
most likely exhibit in their next turn>
Confidence: <a single number between 0 and 1>
\end{lstlisting}
\end{framed}

\subsubsection*{C2: All-In-Context Baseline}
\begin{framed}
\begin{lstlisting}[style=promptstyle]
You are analyzing a real person's verbal 
behavior across a longitudinal study. The 
transcripts below capture how this specific 
individual communicates across multiple 
sessions. Use both the full conversation 
history and the current exchange.

Your task is NOT to predict the exact words 
they will say. Instead, predict their verbal 
behavioral tendency -- the communicative 
intention and function their next utterance 
is most likely to serve, grounded in VRM 
theory.

{VRM_TAXONOMY}

When forming your prediction, consider:
- Which VRM function their next turn will 
  most likely perform
- Stable patterns in how this person has 
  communicated across sessions (e.g., 
  consistent Disclosure, habitual Questions, 
  frequent Confirmation)
- The emotional and relational register they 
  tend to adopt in similar situations
- How the current moment fits or deviates 
  from their longitudinal patterns

Conversation history (prior sessions to 
current context):
{full_context}

Respond with:
Behavioral tendency: <2-3 sentences 
describing the verbal behavior [User] will 
most likely exhibit in their next turn>
Confidence: <a single number between 0 and 1>
\end{lstlisting}
\end{framed}

\subsubsection*{C-NL: Natural Language Baseline}
\begin{framed}
\begin{lstlisting}[style=promptstyle]
You are analyzing a real person's verbal
behavior across a longitudinal study.

{VRM_TAXONOMY}

Below are transcripts from multiple
sessions with this person:

{conversation_data}

Based on these transcripts, identify the
most stable and recurring verbal behavioral
tendencies this person exhibits. For each
tendency:
- Describe it in one plain sentence
  grounded in VRM communicative function
  (e.g., "When discussing personal
  struggles, [User] tends to Disclose
  rather than deflect")
- Support it with a brief direct quote
  from the transcripts

Focus only on patterns that appear
consistently across sessions. Ignore
one-off or situationally unique behaviors.

Respond as a bulleted list:
-- <one-sentence behavioral tendency>
  (e.g., "<supporting quote>")

-- <one-sentence behavioral tendency>
  (e.g., "<supporting quote>")
\end{lstlisting}
\end{framed}
\subsubsection*{C-NL: Natural Language Baseline Prediction}
\begin{framed}
\begin{lstlisting}

[style=promptstyle]
You are predicting a real person's next
verbal act in a conversation.

Based on observations across multiple
sessions, this person shows the following
behavioral tendencies:

-----------------------------------------
{NL_PROFILE}
-----------------------------------------

{VRM_TAXONOMY}

Conversation so far:
{context_text}

[Other] just said: "{other_turn}"

Using the behavioral tendencies above as
evidence for how this person communicates,
describe in 2-3 sentences of plain prose
the verbal behavioral tendency [User] will
most likely exhibit in their next turn --
framed in terms of communicative function
and intent (aligned with VRM categories
above).
Do NOT reference or quote any specific
tendency by name -- use the insights
silently.

Respond with:
Behavioral tendency: <2-3 sentences
describing the verbal behavior [User] will
most likely exhibit in their next turn>
Confidence: <a single number between 0
and 1>
\end{lstlisting}
\end{framed}

\subsubsection*{C3: Situational Reasoning}

\begin{framed}
\begin{lstlisting}[style=promptstyle]
You are predicting a real person's next 
verbal act in a conversation.

The following behavioral patterns are 
ACTIVE right now -- each is grounded in a 
real past example of how this person 
communicated in a similar situation:

-----------------------------------------
{grounded_block}
-----------------------------------------

{VRM_TAXONOMY}

Conversation so far:
{context_text}

Using the past examples above as evidence 
for how this person tends to communicate, 
describe in 2-3 sentences of plain prose 
the verbal behavioral tendency [User] will 
most likely exhibit in their next turn  
framed in terms of communicative function 
and intent (aligned with VRM categories 
above).

Do NOT name or cite any pattern -- use 
the insights silently.

Respond with:
Behavioral tendency: <2-3 sentences 
describing the verbal behavior [User] will 
most likely exhibit in their next turn>
Confidence: <a single number between 0 
and 1>
\end{lstlisting}
\end{framed}

\subsubsection{Behavioral Pattern Extraction \& Pattern-Conditioned Prediction Prompts}

\begin{framed}
\begin{lstlisting}[style=promptstyle]
Based on the following conversation, identify 
recurring verbal behavioral patterns -- stable 
associations between situations and 
communicative acts that characterize how this 
person tends to respond, grounded in VRM theory.

{VRM_TAXONOMY}

Each pattern must follow this format:
  IF [situational antecedent]
  THEN [verbal behavior response]
  BUT NOT when [counter-situation]

{SITUATION_DIMS}

The IF clause is a set of required context 
factors -- ALL must be present to activate. The 
BUT NOT clause is a set of exception factors 
-- if ANY appear, the pattern is suppressed.

For each pattern, provide:
0. logic_formula: The full IF-THEN-EXCEPT rule 
   as a natural language sentence.
   Example: 'If [WITH WHOM: authority figure] 
   + [STATE: user feels defensive], the user 
   tends to deflect; but not when [WITH WHOM: 
   peer].'
1. heuristic_or_pattern: A short label 
   (e.g., 'Deflects-Under-Authority-Pressure')
2. description: One sentence -- 'When __, 
   [User] tends to __ [VRM act], but not 
   when __'.
3. evidence: Quotes from the conversation 
   where the behavior activates, each with:
   - quote: exact text
   - context_label: active IF factors present
   - conversation_id: 4-char ID
   - chain_link: what the quote reveals 
     (trigger, internal state, or verbal act)
4. confidence: Integer 1-10.
   1-2 = never (0.0-0.25)
   3-5 = sometimes (0.25-0.5)
   6-8 = often (0.5-0.75)
   9-10 = always (0.75-1.0)

Only extract patterns with confidence >= 5.

Conversation Data: {conversation}
\end{lstlisting}
\end{framed}

\begin{framed}
\begin{lstlisting}[style=promptstyle]
Given the following behavioral pattern and 
its evidence and counter-evidence, evaluate 
whether the evidence supports the pattern 
and remove evidence that does not. Refine 
the pattern if needed.

{merge_instructions}

First, reason through each piece of evidence 
and counter-evidence (max 200 words). For 
each quote, consider what it reveals about 
the IF-THEN chain: what situational trigger 
activated the behavior, what internal state 
mediated it, and what verbal act resulted.

Then provide a refined pattern:
IF [factor set: WHEN/WHERE/WITH WHOM/STATE]
-- [internal mediating state] --
THEN [verbal act]
BUT NOT when [exception factor set]

If evidence should be removed, list indices 
(0-based) in evidence_indices_to_remove.

Compute probability = {count}, mapped to:
  0.0-0.25 = never
  0.25-0.5 = sometimes
  0.5-0.75 = often
  0.75-1.0 = always

pattern:
{hypothesis}

Evidence (indexed):
{indexed_evidence}

Counter-evidence:
{counterfactuals}
\end{lstlisting}
\end{framed}

\begin{framed}
\begin{lstlisting}[style=promptstyle]
You are deciding whether [Other]'s message 
will trigger a specific behavioral pattern 
in [User].

Each pattern has the form:
  IF [situational antecedent]
  THEN [verbal behavior response]
  BUT NOT when [counter-situation]

{SITUATION_DIMS}

Firing works as set membership:
- The IF clause activates only when ALL required 
  factors are present
- The BUT NOT clause suppresses activation if 
  ANY of its factors appear
- A partial IF match does NOT activate

For each pattern, work through:
Step 1 -- EXTRACT: List active context 
  factors labeled by dimension.
Step 2 -- IF SUBSET CHECK: Are ALL IF 
  factors present? If any missing, do NOT 
  activate.
Step 3 -- BUT NOT CHECK: Does the factor 
  set contain ANY BUT NOT factors? If yes, 
  suppress.
Step 4 -- THEN-BEHAVIOR LIKELY: Is the 
  THEN-behavior the most likely next verbal 
  act, not just possible?
Step 5 -- NOT AMBIGUOUS: Could [User] easily 
  give a completely different response? If 
  yes, do NOT activate.

behavioral patterns:
-----------------------------------------
{pattern_block}
-----------------------------------------

[Other]'s message: "{other_turn}"

List ONLY patterns that pass ALL steps 
with confidence >= 8:
Pattern: <name>
Active factor set: <factors with labels>
IF subset match: <confirm all IF factors>
BUT NOT clear: <confirm no BUT NOT factors>
Trigger active: <why THEN-behavior is most 
  likely>
Confidence: <8-10>

Default answer: NONE.
\end{lstlisting}
\end{framed}

\subsubsection{Judge Prompts}
\label{appx:judge_prompt}

\begin{framed}
\begin{lstlisting}[style=promptstyle]
You are evaluating how well a predicted 
verbal behavioral tendency matches what a 
person actually said. Work through three 
dimensions step by step, then give a single 
aggregated score from 0 to 1.

Predicted tendency: "{predicted}"
Actual utterance:   "{actual}"

{VRM_TAXONOMY}

---

DIMENSION 1 -- Pragmatic Function Alignment
Does the prediction capture the same verbal 
intention as the actual utterance, independent 
of surface wording? Focus on the VRM 
communicative function -- the why, not the 
what.

Examples:
- Predicted deflection / actual withdraws 
  and changes subject -- HIGH
- Predicted clarifying question / actual 
  withdraws -- LOW (wrong VRM function)
- Predicted agreement / actual briefly 
  confirms and moves on -- MEDIUM-HIGH

Step 1a: Identify the VRM function the 
actual utterance performs.
Step 1b: Identify the VRM function the 
prediction describes.
Step 1c: How closely do they match?
D1 reasoning: <your reasoning>

---

DIMENSION 2 -- Behavioral Specificity
Does the prediction match the level of 
specificity of the actual utterance? A 
prediction that could apply to almost any 
situation scores low.

Examples:
- Predicted 'minimizes distress and 
  redirects to logistics' / actual does 
  exactly this -- HIGH
- Predicted 'may express concern' / actual 
  minimizes and redirects -- LOW (too 
  generic)
- Predicted 'will push back' / actual 
  disagrees on specific point -- MEDIUM

Step 2a: How specific is the actual 
utterance's behavior?
Step 2b: Is the prediction equally specific?
Step 2c: Would this prediction apply to 
many other turns, or is it tuned to this one?
D2 reasoning: <your reasoning>

---

DIMENSION 3 -- Compositional Alignment
When the actual utterance performs multiple 
functions, does the prediction capture them 
and reflect their relative weight?

Examples:
- Actual: Discloses then redirects with 
  Advisement; predicted captures both -- HIGH
- Same actual; predicted captures only 
  Disclosure -- MEDIUM
- Same actual; predicted captures only 
  Advisement -- LOW

Step 3a: How many functions does the actual 
utterance perform?
Step 3b: Does the prediction account for 
all of them?
Step 3c: Is the weighting proportional?
D3 reasoning: <your reasoning>

---

Assign a single aggregated score weighting 
dimensions by relevance to this case. Use 
the full range from 0 to 1.

Respond in this exact format:
D1 reasoning: <reasoning>
D2 reasoning: <reasoning>
D3 reasoning: <reasoning>
Score: <a single number between 0 and 1>
\end{lstlisting}
\end{framed}

\subsection{Verbal Response Mode (VRM)}
\begin{table}[h]
\centering
\caption{Verbal Response Mode (VRM) taxonomy. 
Each mode classifies the communicative function 
an utterance serves}
\label{tab:vrm}
\begin{tabular}{p{2.2cm}p{5.8cm}}
\toprule
\textbf{Mode} & \textbf{Description} \\
\midrule
Disclosure & Revealing one's own feelings, 
thoughts, or personal experiences; making 
the inner world visible. \\
\addlinespace
Edification & Conveying information about 
the world or a situation in an objective 
or descriptive way. \\
\addlinespace
Advisement & Influencing what the other 
person does or thinks; giving instructions, 
suggestions, or requests. \\
\addlinespace
Confirmation & Signaling understanding, 
agreement, or support; acknowledging what 
the other said in a validating way. \\
\addlinespace
Question & Explicitly requesting information 
or clarification; creating an expectation 
that something will be provided. \\
\addlinespace
Reflection & Feeding back the other person's 
own experience; paraphrasing or mirroring 
what they communicated. \\
\addlinespace
Interpretation & Going beyond what was said 
to offer an explanation, inference, or 
reframing of the other person's experience. \\
\addlinespace
Acknowledgement & Responding in a primarily 
social or ritualistic way; brief responses 
that maintain conversational flow. \\
\bottomrule
\end{tabular}
\end{table}

\section{Corpus Statistics}

\begin{figure*}
    \centering
    \includegraphics[width=0.99\linewidth]{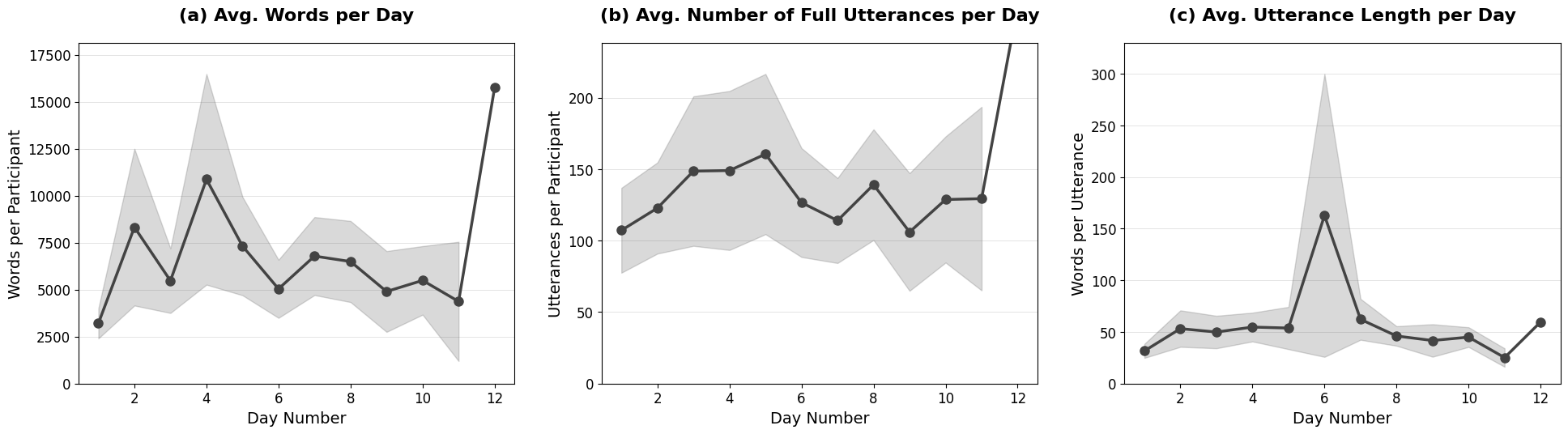}
    \includegraphics[width=0.99\linewidth]{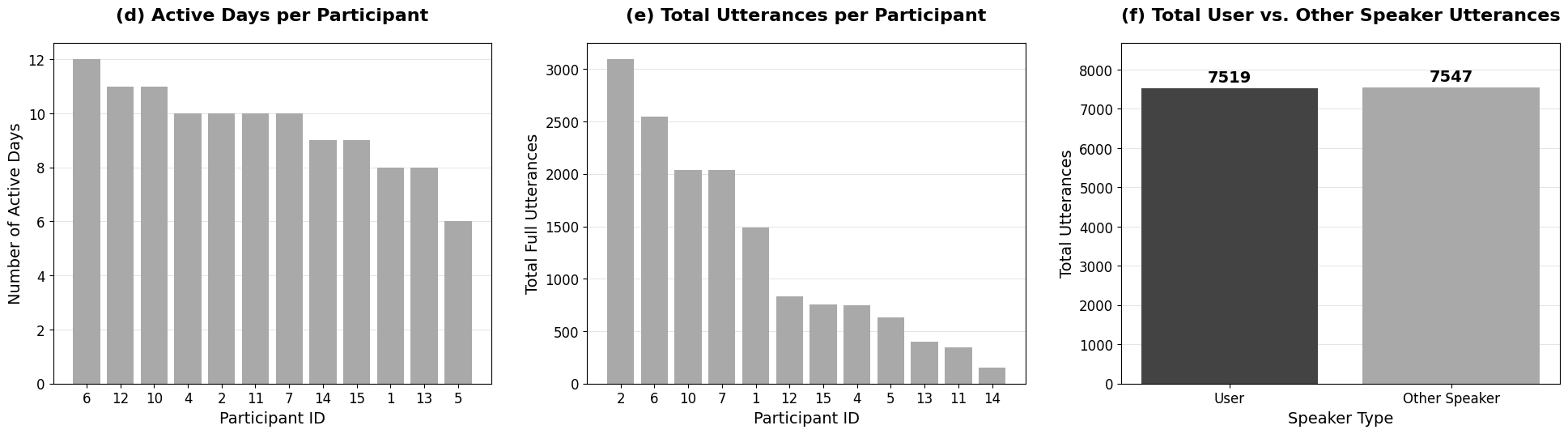}
    \caption{Watch transcription activity for {completed participants ($n=14$)} over the {6-12 day} experiment period. {(a) The average words transcribed within each participant's data for each day. (b) The average number of full utterances for each participant's data for each day including both the user and other speakers. (c) The average word length for each full utterance for each participant's data for each day. (d) The number of days where the watch was actively transcribing for each participant. (e) The total number of full utterances across all participants. (f) A comparison of the total number of full utterances for transcriptions across all participants' data tagged as the `user' and as `other'.}}
    \label{fig:desc}    
    \Description{Six plots summarizing watch transcription activity for 14 completed participants over a 6--12 day study period. 
(a) A line plot showing average words transcribed per participant per day, with noticeable day-to-day variability and a late spike. 
(b) A line plot of the average number of full utterances per participant per day, generally stable with an increase on the final day. 
(c) A line plot of average utterance length in words per day, mostly consistent with a single pronounced peak mid-study. 
(d) A bar chart showing the number of active transcription days per participant, ranging from 6 to 12 days. 
(e) A bar chart of total full utterances per participant, showing substantial variation across participants. 
(f) A bar chart comparing total utterances labeled as user versus other speaker, showing similar overall counts.}
\label{corpus}
\end{figure*}

\end{document}